\documentclass[aip,reprint,pop]{revtex4-1}

\usepackage{graphicx}
\usepackage{subfig}
\usepackage{epstopdf}
\usepackage{amsmath}
\usepackage{dcolumn}
\usepackage[version=4]{mhchem} 
\usepackage{bm}
\usepackage[colorlinks=true, linkcolor=blue]{hyperref}

\begin{document}

\title{Opportunities for plasma separation techniques in rare earth elements recycling}
\author{R. Gueroult}
\affiliation{Laplace, Universit\'{e} de Toulouse, CNRS, INPT, UPS, 31062 Toulouse, France}
\author{J.~M.~Rax}
\affiliation{Universit\'{e} de Paris XI - Ecole Polytechnique, LOA-ENSTA-CNRS, 91128 Palaiseau, France}
\author{N. J. Fisch}
\affiliation{Princeton Plasma Physics Laboratory, Princeton University, Princeton, NJ 08543, USA}

\begin{abstract}
Rare earth elements recycling has been proposed to alleviate supply risks and market volatility. In this context, the potential of a new recycling pathway, namely plasma mass separation, is uncovered through the example of nedodymium - iron - boron magnets recycling. Plasma mass separation is shown to address some of the shortcomings of existing rare earth elements recycling pathways, in particular detrimental environmental effects. A simplified mass separation model suggests that plasma separation performances could compare favourably with existing recycling options. 
In addition, simple energetic considerations of plasma processing suggest that the cost of these techniques may not be prohibitive, particularly considering that energy costs from solar may become significantly cheaper. Further investigation and experimental demonstration of plasma separation techniques should permit asserting the potential of these techniques against other recycling techniques currently under development.  
\end{abstract}

\maketitle

\section{Introduction}
Owing to their unique ferromagnetism, superconductivity and luminescence properties,  rare earth elements (REEs) are key components in a large number of technologies. Although REEs have long been used in mature markets, such as catalysts, glassmaking and lighting, REEs are now also sought after by emerging, high-growth markets, such as permanent magnets and battery alloys~\cite{Goonan2011,Zepf2013,Brumme2014}. Quantitatively, the global REEs mining expanded on average by $7\%$ annually between 1990 and 2006~\cite{Goonan2011}, while REEs demand for permanent magnet manufacturing alone grew by $280\%$ between $2000$ and $2007$ (equivalent to $16\%$ annually)~\cite{Yang2016}. Looking ahead, the globalization of low-carbon energy systems, and in particular of wind turbines and electric cars, is projected to yield demand growth of $700\%$ and $2600\%$ over the next $25$~years for respectively neodymium (Nd) and dysprosium (Dy)~\cite{Alonso2012} by virtue of their unmatched performances for high strength, high temperature magnets~\cite{Jacoby2013}.


On the supply side, REEs production is extremely unequally distributed worldwide. A single country - China - is responsible for over $80$ percent of the current global mining production (and even for as much as $97\%$ up until $2010$~\cite{BGS2013,Chakhmouradian2012}), and controls over $50\%$ of worldwide mineral reserves~\cite{Hatch2012,USGS2013,Humphries2013,USGS2015,USGS2017}. A direct consequence of this near-monopoly situation is a very vulnerable market and a high volatility of REEs price. This market volatility led to the price spike of $2011$, during which the price of some REEs experienced a 10-fold increase over a few months~\cite{DOE2011,Hatch2012}. The uncertainty with respect to price, availability, and quality of raw materials is a serious concern for many industries
in countries that are almost $100\%$ import-reliant, such as the USA and multiple E.~U. countries. This situation led panels and governmental agencies to place REEs, in particular Nd, yttrium (Y), Dy, europium (Eu) and terbium (Tb), on their \emph{critical} raw elements list~\cite{Moss2011,DOE2011}. Furthermore,  the main consumer countries began implementing mineral strategies to minimise their vulnerability to the supply of REEs~\cite{NRMRL2012,EC2013}.

One strategy is supply diversification~\cite{Golev2014,Bartekova2016}. This option has led to the start-up or reviving of exploration projects and mining production around the world to alleviate supply risks~\cite{BGS2011,Mariano2012,Sarapaa2013,Goodenough2016,Machacek2016}. The trigger effect of the $2011$ price spike in the geographical diversification of REEs mining is hardly debatable. As a matter of fact, China had $97\%$ of the REEs market share up until $2010$~\cite{BGS2013}, whereas about a half dozen countries produced REEs in $2015$, with Australia, India, the USA, and Russia combining for about $17\%$~of the global production~\cite{BGS2016, USGS2017}. Although mining expansion has obvious benefits from a market stability perspective, REEs mining comes at a significant environmental cost~\cite{Humsa2015,Charalampides2016,Lee2016,Browning2016,Huang2016}. In addition, health hazards associated with REEs mining are only beginning to receive attention~\cite{Li2013,Rim2013,Pagano2015,Rim2016}. Although it has been suggested that the environmental impact of mining could be reduced if subjected to stricter environmental legislation~\cite{Schreiber2016}, it remains unclear whether REEs mining will be societally accepted in european countries, as exemplified by the ongoing debate in Denmark on REEs mining in Greenland~\cite{Rosen2016}. Finally, mining expansion does not address the so-called \emph{balance} problem~\cite{Falconnet1985,Binnemans2015}, which stems from the fact that natural REEs abundance in ores does not match the market demand of each individual REE.

Another option to address the supply-risks while limiting environmental damage and mitigating the balance problem is recycling~\cite{Bloodworth2013,Tsamis2014,Verrax2015,Bartekova2016}. Correlation between recycling rate and market price stabilization has notably been observed previously for cobalt and platinum~\cite{Darcy2013,Verrax2015,Bandara2015}. Despite these strong incentives, less than $1\%$ of REEs were recycled in $2011$~\cite{Graedel2011,Reck2012,Binnemans2013}. Various reasons have been given to explain REEs low recycling level, such as the lack of effective collection systems, the difficultly to extract REEs from scrap, or the relatively low prices prior to 2011. However, one of the main impediment to the development of REEs recycling appears to be the low REEs content of most end-products~\cite{Golev2014}. Mass content of REEs for most applications is lower than $5$~g/kg ($0.5\%$), and as low as $0.5$~g/kg ($0.05\%$) in LEDs~\cite{Chancerel2013}. One exception is neodymium - iron - boron (\ce{Nd2Fe14B}, or NdFeB for short) permanent magnets, for which the REEs mass content can be as large as $30\%$. 

Interestingly, NdFeB magnets currently happen to dominate the permanent magnets market thanks to their superior energy product~\cite{Jacoby2013}. Small NdFeB magnets are used extensively in consumer products, such as hard disk drives and loudspeakers, while large NdFeB magnets are increasingly used in electric vehicles and windmill turbines~\cite{Binnemans2013}. Although large magnets can be efficiently disassembled to be recycled, disassembling of electronic goods is key in making REEs recovery from these products attractive~\cite{Tsamis2014}. Indeed, without efficient pre-processing, rare earth magnets in electronic goods are shredded along with waste electronics and electrical equipment, decreasing in turn significantly the mass content of REEs. 


In this paper, we illustrate the potential of plasma mass separation techniques for rare earth elements recycling through the example of NdFeB magnets.
The two main rare earth recycling pathways are first briefly reviewed, and
the general features and intrinsic advantages of plasma separation are introduced. 
The potential of plasma mass separation for NdFeB magnets recycling is then discussed, 
and a preliminary cost estimate is derived. 
Finally, concept improvements are suggested. 

\section{Options for rare earth magnets recycling}

\subsection{Current trends}
\label{Sec:II}

Owing to their large REEs content and widespread use, \ce{NdFeB} magnets hold promise for REEs urban mining. In light of this opportunity, \ce{NdFeB} magnets recycling has sparked significant interest in the last few years~(see, \emph{e.~g.}, Refs.~\cite{Rademaker2013,Rabatho2013,Sprecher2014,Onal2015,Bogart2015,Bandara2016,Bian2016}).

The two main recycling pathways for \ce{NdFeB} magnets considered to date are hydro-metallurgy and pyro-metallurgy~(see, \emph{e.~g.}, review articles\cite{Binnemans2013,Takeda2014,Firdaus2016,Yang2016}). Hydro-metallurgical recycling typically involves processes very similar to those used for REEs extraction from primary ores. These processes rely heavily on the use of chemicals, including strong mineral acids. This generally leads to the production of large volumes of liquid wastes and a significant environmental footprint. Note however that new separation schemes which coud alleviate some of these limitations are currently under development~\cite{Bogart2015}. 

On the other hand, pyro-metallurgical or high-temperature processing is believed to be more environmentally friendly, but at the expense of a greater energy input. Yet, some high-temperature processes require chemicals, and produce complex wastes~\cite{Firdaus2016}. A promising recent development in high temperature recycling is the vacuum induction melting - magnetic separation (VIM-HMS) process~\cite{Bian2016}, which requires no toxic chemicals and produces minimal solid waste. However, this process might be challenged by complex compositions, in particular other transition metals such as copper and nickel~\cite{Yang2016}. 

Finally, new recycling process relying on REEs absorption on bio-materials such as salmon milt~\cite{Takahashi2014} and bacteria~\cite{Bonificio2016,Park2016} have recently been proposed.

\subsection{Plasma separation}
\label{Sec:III}

Yet another possibility for REEs recycling might be plasma separation. By operating on dissociated molecules, plasma separation can be used where chemical techniques are challenged. Applications for which these capabilities could prove extremely valuable include nuclear legacy waste disposal~\cite{Siciliano1993,Freeman2003,Gueroult2015} and spent fuel reprocessing~\cite{Zhiltsov2006,Gueroult2014a,Timofeev2014}. By accommodating straightforwardly complex chemical compositions, plasma separation could in principle handle coatings, additives and contaminants which are typically found in permanent magnets~\cite{Firdaus2016}. In addition, plasma separation could handle equally sintered magnets or resin bonded magnets. Furthermore, plasma separation does not require chemicals, nor creates additional waste streams. By virtue of these properties, plasma separation could in principle offer an environmentally friendly pathway for REEs recycling. However, as it will be shown later, the energy input required to separate products scale linearly with the number of atoms in the feed for plasma separation. Plasma separation is therefore expected to be most attractive for concentrated feeds such as large magnets, and less for diluted feeds such as waste electronics.
 
Although plasma separation can be envisioned in many ways, it essentially boils down to three core technologies. First, the input stream has to be turned into a plasma. By heating material to very high temperatures, the bonds that commonly form chemical substances are broken.  Upon further heating, atoms are ionized, leaving individual ions and electrons, \emph{i~e.} a plasma. A variety of well established techniques can be used for plasma formation, including for example laser ablation~\cite{Boulmer1993}, arc-discharges~\cite{Brown2005}, or dust injection~\cite{Tanaka2007}. Then, once a plasma is formed, the second technology consists in preferentially extracting specific elements from the bulk. This is where we believe plasmas are unique, as discussed next. Finally, the third technology corresponds to the collection of separated elements. This can be achieved by depositing elements on surfaces, or through recombination in volume.

For separation purposes, plasmas stand out from other states of matter (like gas and liquid) owing to the ability to control and manipulate particles through their electric charge. It then becomes possible to combine electromagnetic forces with other forces to produce differential effects. By harnessing these effects, elements can be separated. For example, a rotating plasma produces mass separation~\cite{Bonnevier1966,Bonnevier1971,Lehnert1971,Oneil1981} due to the combined effects of electromagnetic and centrifugal forces. This mechanism was first put to work in plasma centrifuges~\cite{Lehnert1973,Krishnan1981,DelBosco1991} to separate isotopes~\cite{James1976,Prasad1987,Grossman1991} in the 1980s. It is worth noting here that although plasma centrifuges are conceptually similar to liquid or gas centrifuges, they differ in that rotation is produced through electromagnetic effects, which allows to operate plasma centrifuges at much larger rotation speeds than their liquid and gaseous counterparts. Since the separation power of a centrifuge depends on the square of the rotation speed~\cite{Fetterman2011b}, higher separation are achievable in a single unit. 

In contrast with isotope separation, which involves separating small quantities of elements with very small mass differences, some applications (\emph{e.~g} nuclear legacy waste disposal and nuclear spent fuel reprocessing) require separating elements with large mass differences at high-throughput. To address this need, a suite of new rotating plasma configurations expanding on the plasma centrifuge concepts has recently been theoretically studied~\cite{Ohkawa2002,Fetterman2011,Gueroult2012a,Gueroult2014,Rax2016,Ochs2017,Gueroult2017c}. Beyond rotating configurations, which are obvious candidates, a wide variety of differential mechanisms and configurations can in principle be used to discriminate elements based on mass in plasmas. For example, plasma optical separators~\cite{Morozov2005,Bardakov2010}, crossed-field plasma separators~\cite{Smirnov2013}, and plasma filters based on differential magnetic~\cite{Timofeev2000} and collisionality gradient~\cite{Ochs2017a} drifts, and gyro-radius effects~\cite{Babichev2014} have all been suggested to separate elements based on atomic mass.


It is important to note here that although plasma centrifuges have been demonstrated experimentally (see, \emph{e.~g.}, Refs.~\cite{Krishnan1983,Hirshfield1989}), experimentation on high-throughput separation of elements with large mass differences is still in a relatively primitive stage of development, and limited thus far to laboratory scale~\cite{Bardakov2014,Paperny2015,Vorona2015,Gueroult2016a}. Full scale demonstration of the entire plasma separation process thus remains to be made. 

\section{Results and Discussion}

\subsection{NdFeB magnets plasma recycling}
\label{Sec:IV}

Typical composition of Rare Earth magnets (REMs) not only contains Nd, Fe and B, but also some mass-percent of Dy to increase the magnets maximum operating temperature, as well as various amount of Al, Co, Ni and Nb used to improve mechanical properties~\cite{Firdaus2016}. In addition, contaminants like C, Ca, N, Si and O are found in REMs waste. 

\begin{figure}
\includegraphics{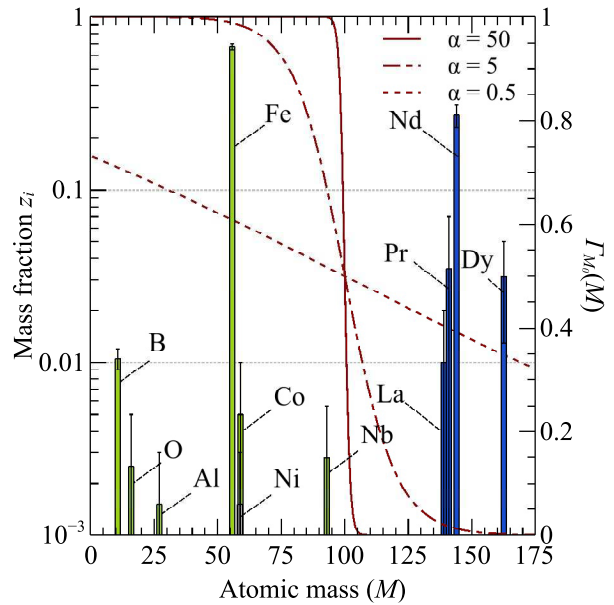}
\caption{Typical composition of Rare Earth permanent magnet waste (from Ref.~\cite{Firdaus2016}) (left vertical axis, log-scale) along with the idealized plasma mass filter function, ${\Gamma\!}_{M_0}(M)$ defined in Eq.~(\ref{Eq:idealized_filter}) (right vertical axis, linear-scale). $M_0=100$~amu and $\alpha=50$ (case $\mathcal{A}$), $\alpha=5$ (case $\mathcal{B}$) and $\alpha=0.5$ (case $\mathcal{C}$). Elements making less than $0.1\%$ in mass ($z_i\leq10^{-3}$) are not shown.}
\label{Fig:Composition}
\end{figure}

The presence of these additional elements in the input stream represents a challenge for high-temperature recycling~\cite{Firdaus2016}. In contrast, the potential of plasma mass separation for REMs recycling becomes evident when plotting REMs waste composition by mass as a function of its constituents atomic mass. As illustrated in Fig.~\ref{Fig:Composition}, REMs waste in its elemental form breaks down into two groups. On the one hand, one finds a group of light elements ($m_i\leq 92.9$~amu, in light green) made of Fe, B, coating, additives and contaminants. Iron (Fe) makes over $96\%$ of the mass of these light elements. On the other hand, the group of heavy elements ($m_i\geq 138.9$~amu, in blue) is exclusively made of REEs. Plasma mass filtration could in principle efficiently separate the heavy group, that is to say REEs, from the light one. The same mass filtering process will operate equally on resin-bonded magnets. Indeed, a threshold mass around $110$~amu will collect resin elements (mostly carbon, hydrogen and oxygen) along with light elements due to their low atomic mass, effectively separating REEs from non REEs.

\begin{figure}
\includegraphics{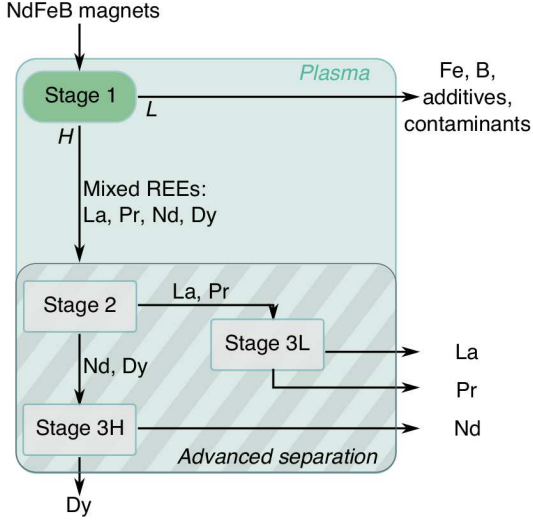}
\caption{Flowchart envisioned for NdFeB magnets plasma recycling. Stage 1 separates mixed REEs (stream $H$) from Fe, B and additives and contaminants (stream $L$), and is the main object of this study. The lower part (hatched region), referred to as advanced separation, illustrates how the mixed REEs stream produced by stage 1 could be further processed to extract individual REEs.  }
\label{Fig:FlowChart}
\end{figure}

In the once-through process proposed above,  REEs are separated from all non-REEs elements, but not from one another. In the case of an input stream containing multiple REEs such as the one plotted in Fig.~\ref{Fig:Composition}, the output therefore consists of a mixture of REEs. This process corresponds to stage $1$ in the plasma flowchart depicted in Fig.~\ref{Fig:FlowChart}, and matches the output of a variety of separation processes recently investigated, for which individual REEs are not separated (see, \emph{e.~g.}, Refs~\cite{Onal2015,Bandara2016,Bian2016}). Note that although this will not be the primary focus of this study, it is in principle possible to extract individual REEs from one another via plasma mass separation as illustrated by the lower part of Fig.~\ref{Fig:FlowChart}. This possibility is briefly discussed at the end of the this section. 


As indicated in the previous section, 
high throughput plasma mass separation can be envisioned in multiple ways. Rather than attempting to capture the subtle differences between these configurations and separation mechanisms, we choose to discuss the potential of plasma separation for \ce{NdFeB} magnets recycling through their common characteristic, that is to say mass discrimination. For this purpose, let us define here an analytical filter function of the form 
\begin{equation}
{\Gamma\!}_{M_0}(M) = \frac{1+\tanh(\alpha[1-M/M_0])}{2},
\label{Eq:idealized_filter}
\end{equation}
with $\alpha$ a real number representing the inverse of the filter width, $M$ the element atomic mass and $M_0$ the filter threshold mass. Three different cases $\mathcal{A}$, $\mathcal{B}$ and $\mathcal{C}$, obtained for $M_0 = 100$~amu and respectively $\alpha = 50$, $5$ and $0.5$, are plotted in Fig.~\ref{Fig:Composition}. Case $\mathcal{A}$ represents a very narrow, almost  ideal filter, for which ${\Gamma\!}_{M_0}(M)$ tends to a step function. Case $\mathcal{B}$ is significantly broader, but the filter width remains comparable to the mass gap observed between non-REEs and REEs. Finally, for case $\mathcal{C}$, ${\Gamma\!}_{M_0}(M)$ is nearly linear over the mass range considered. Introducing $x_i$, $y_i$ and $z_i$ the mass fraction of element $i$ of mass $M_i$ in respectively the heavy, light and input stream, one gets 
\begin{equation}
x_i = \frac{z_i\left[1-{\Gamma\!}_{M_0}(M_i)\right]}{\sum\limits_{j}z_j\left[1-{\Gamma\!}_{M_0}(M_j)\right]}\quad \textrm{and} \quad y_i = \frac{z_i{\Gamma\!}_{M_0}(M_i)}{\sum\limits_{j}z_j{\Gamma\!}_{M_0}(M_j)}
\end{equation}
where $j$ runs over all elements, including $i$. We then define the separation factor
\begin{equation}
\beta_i = \frac{x_i(1-z_i)}{z_i(1-x_i)}
\end{equation}
and the extraction efficiency
\begin{equation}
r_i = \frac{\theta x_i}{z_i},
\end{equation}
where $\theta$ is the cut, that is to say the ratio between the product (here called heavy) and input flow.

\begin{figure}
\includegraphics{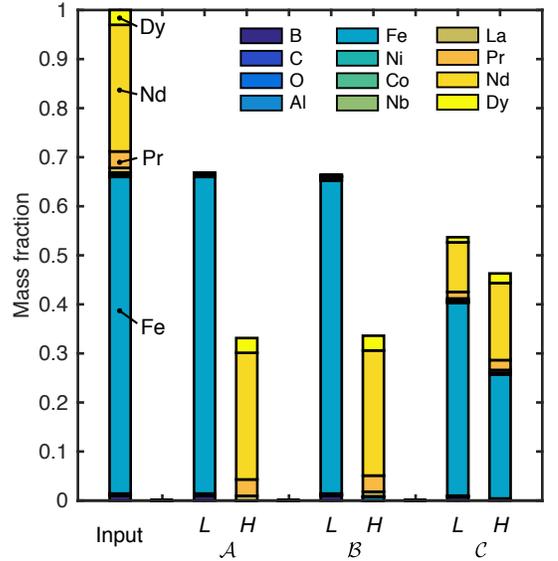}
\caption{Input stream (leftmost stack) along with mass fraction in each separated stream $L$ (left stack) and $H$ (right stack) for each of the three cases $\mathcal{A}$, $\mathcal{B}$ and $\mathcal{C}$. The REEs purity in the heavy stream ($H$), ${x_{\textrm{RE}}}$, is $1$, $0.97$ and $0.44$, respectively.}
\label{Fig:FilteredComposition}
\end{figure}

The composition of the two streams produced by this idealized mass filter (light elements stream $L$ and heavy elements stream $H$) is plotted in Fig.~\ref{Fig:FilteredComposition} for each of the three cases, along with the input stream composition. 

For cases $\mathcal{A}$ and $\mathcal{B}$, the mass fraction of REEs in the heavy stream ${x_{\textrm{RE}}}$ is respectively $1$ and $0.97$, indicating a very good separation of REEs from both Fe and B, as well as coatings, additives and contaminants. In terms of separation factor, one gets $\beta_{\textrm{RE}} = 3~10^5$ and $76$ for the same conditions. The REEs extraction efficiency is $r_{\textrm{RE}} = 1$ and $r_{\textrm{RE}}=0.99$ for $\alpha=50$ (case $\mathcal{A}$) and $\alpha=5$ (case $\mathcal{B}$), respectively. It is worth noting that such REEs purity levels and extraction efficiencies compare favourably with existing recycling processes~\cite{Takahashi2014,Onal2015,Bandara2016,Bogart2015,Bian2016}. 

In contrast, for $\alpha = 0.5$ (case $\mathcal{C}$), REEs mass fraction in the heavy stream is only $0.44$, the extraction efficiency is $0.61$, and the separation factor is $1.57$ (see Tab.~\ref{Ref:Tab1}). This poor purity level stems from the large fraction of Fe collected in the heavy elements stream owing to the very large width of the filter, as seen in Fig.~\ref{Fig:Composition}. Another consequence of the large filter width is the presence of a significant amount of REEs in the light elements stream, which in turn yields low extraction efficiency.  These results suggest that a filter with such a large width is unlikely to be appropriate for REEs recycling, or at least not in a single step. 

\begin{table}
\begin{center}
\begin{tabular}{c | c | c | c}
 & Case $\mathcal{A}$ & Case $\mathcal{B}$  & Case $\mathcal{C}$\\ 
 \hline
Inverse filter width $\alpha$ & $50$ & $5$ & $0.5$\\
REEs purity in output stream $x_\textrm{RE}$ & $1$ & $0.97$ & $0.44$\\
REEs separation factor $\beta_\textrm{RE}$ & $3\times 10^5$ & $76$ & $1.57$\\
REEs extraction efficiency $r_\textrm{RE}$ & $1$ & $0.99$ & $0.61$\\
Cut $\theta$ & $0.33$ & $0.34$ & $0.46$ \\
\end{tabular}
\caption{Idealized filter parameters and REEs separation performances. For all three cases, $M_0 = 100$~amu. }
\label{Ref:Tab1}
\end{center}
\end{table}

\begin{figure}
\includegraphics{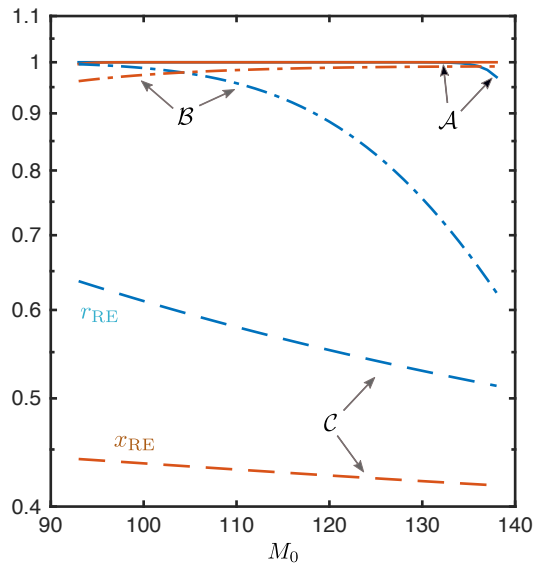}
\caption{Extraction efficiency $r_{\textrm{RE}}$ (blue) and REEs purity in the heavy stream ${x_{\textrm{RE}}}$ (red) as a function of the filter threshold mass $M_0$ for different filter widths ($\alpha = 50$ - solid, $\alpha = 5$ - dash-dotted and $\alpha = 0.5$ - dashed). The threshold mass is varied between the heaviest non REEs (Nb) and the lightest REEs (La) typically found in REMs. }
\label{Fig:Alpha_M_0_dependency}
\end{figure}

Further to the sensitivity to the filter width studied thus far, the filter mass threshold can be tuned to optimize separation.  For an intermediate filter width, there exists a trade-off between the extraction efficiency and the REEs purity, as illustrated by case $\mathcal{B}$ in Fig.~\ref{Fig:Alpha_M_0_dependency}. A high mass-threshold will maximize purity at the expense of the extraction efficiency, and vice-versa. For a filter width narrower than the mass gap between non-REEs and REEs (case $\mathcal{A}$), little effect is observed other than for a decrease in extraction efficiency as $M_0$ approaches the mass of the lightest REEs. Finally, for a large filter width (case $\mathcal{C}$), both $r_\textrm{RE}$ and $x_\textrm{RE}$ decrease with $M_0$. The result that $x_{\textrm{RE}}$ decreases with $M_0$ for small $\alpha$ might seem counter-intuitive at first. However, expanding the filter function ${\Gamma\!}_{M_0}(M)$ for small $\alpha$,
\begin{equation}
{\Gamma\!}_{M_0}(M) = \frac{1+\alpha}{2}-\frac{\alpha}{2M_0}M+\mathcal{O}(\alpha),
\end{equation}
shows that the filter slope for $\delta M = |M-M_0|\ll\alpha^{-1}$ is in this limit inversely proportional to $M_0$, which explains the decrease of $x_{\textrm{RE}}$ with $M_0$. 

\begin{figure}
\includegraphics{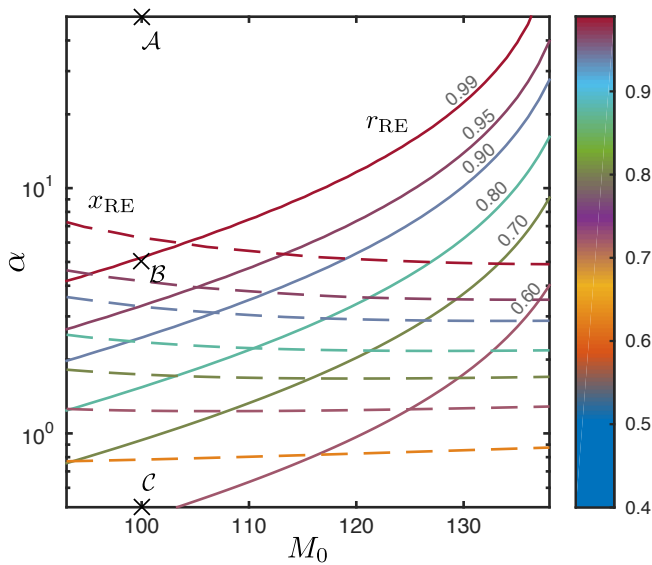}
\caption{Contours of the REEs extraction efficiency $r_{\textrm{RE}}$ (solid lines) and the REEs purity in the heavy stream ${x_{\textrm{RE}}}$ (dashed lines) over the $(M_0,\alpha)$ space. Parameters corresponding to cases $\mathcal{A}$, $\mathcal{B}$ and $\mathcal{C}$ are indicated with black crosses. }
\label{Fig:Full_alpha_M_0}
\end{figure}

For completeness, the evolution of $x_{\textrm{RE}}$ and $r_{\textrm{RE}}$ over the whole range of $(M_0, \alpha)$ considered is plotted in Fig.~\ref{Fig:Full_alpha_M_0}. This result shows that $M_0$ has little influence over $x_{\textrm{RE}}$, which indicates that purity level is mostly controlled by the filter width. In addition, it confirms the sign of $\partial x_{\textrm{RE}}/\partial M_0$ changes from positive to negative as the filter width grows. In contrast, $r_{\textrm{RE}}$ decreases with $M_0$ for all $\alpha$. In addition, Fig.~\ref{Fig:Full_alpha_M_0} reveals that high efficiency separation such that both $x_{\textrm{RE}}$ and $r_{\textrm{RE}}$ are greater than $99\%$ can be obtained for parameters less demanding than those corresponding to case $\mathcal{A}$, for example $(M_0,\alpha) = (104,6)$. A relaxed criteria of $95\%$ can be achieved for $(M_0,\alpha) = (105,4)$. Finally, $\alpha=2$ for $M_0 = 94$ is enough to produce a $99\%$ pure REEs stream ($x_{\textrm{RE}}\geq 0.99$) if tolerating a lower extraction efficiency $r_{\textrm{RE}} \sim 0.75$.

Although very rudimentary, this simplified filter model provides valuable pointers to what a plasma mass filter function should be in order to be applicable and promising for REEs recycling. However, since experimental filter functions for the proposed plasma filter concepts have not yet been produced to support this model, conclusions drawn from this model should be regarded only as approximate. Still, on the basis of successful plasma isotope separation experiments, it stands to reason that filter functions approximating Eq.~(\ref{Eq:idealized_filter}) could be produced, at least at moderate throughput. If so, the results obtained above will hold true, at least qualitatively. In particular, we note that a filter with a width of the order of the mass gap between non-REEs and REEs (case $\mathcal{B}$) should still provide high performances for \ce{NdFeB} magnets recycling. 

It is worth emphasizing here that the filter function of a real plasma mass filter will depend on very many parameters. To name a few, plasma density, electron and ion temperatures, neutral background density and drift velocities will most likely all play a role on the shape of the filter function. Producing a specific filter function will consequently require controlling simultaneously a rather large numbers of plasma parameters. Yet, this complexity is in principle balanced by the possibility to fine tune the filter function thanks to the large number of control knobs. One desirable feature would for example be asymmetry, that is to say 
\begin{equation}
{\Gamma\!}_{M_0}(M_0+M)+{\Gamma\!}_{M_0}(M_0-M)\neq 1.
\end{equation}



\subsection{Preliminary cost assessment}
\label{Sec:V}

Although a detailed cost calculation is beyond the scope of this work, and will require in-depth knowledge of the selected plasma filter concept, a first estimate for the cost of \ce{NdFeB} magnets (REMs) plasma processing can be obtained on account of the core technologies of a generic plasma filter which were introduced in the previous section. 
To this end, the energy cost associated respectively with input stream evaporation and with plasma production and maintenance are evaluated in this section. 
 
Since plasma processing involves breaking the input feed into elemental form, its is useful to recall how many atoms need to be processed per kg of input feed. One kilogram of pure \ce{Nd2Fe14B} alloy is made of $\mathcal{N} = 9.42~10^{24}$~atoms, leading to a Nd mass fraction $z_{\textrm{Nd}} = 0.27$. The number of atoms in one kilogram of pure \ce{Dy2Fe14B} alloy differs from $\mathcal{N}$ by only $3\%$, so that the presence of small fraction of other REEs will be neglected in first approximation. For the same reason, the presence in REMs of small mass fraction of additives and contaminants, or variations in cast alloys composition (for example Nd$_{2.6}$Fe$_{13}$B$_{1.4}$, or \ce{Nd15Fe77B8}~\cite{Firdaus2016}), are neglected. With this information in hand, it is now possible to evaluate the energy cost of \ce{NdFeB} plasma processing.

The energy input required to turn solid material into gas is the sum of the enthalpy change corresponding to heating the material from room temperature $T_0$ to fusion temperature $T_f$ at a given pressure, the latent heat of fusion $\mathcal{L}_F$, the enthalpy change corresponding to heating the material from fusion temperature to boiling temperature $T_b$ at a given pressure, and the latent energy of vaporization $\mathcal{L}_V$. The ratio $\mathcal{L}_F/T_f$ or entropy of fusion is roughly constant for metals and of the order of $R = 8.314$~J.mol$^{-1}$.K$^{-1}$ (Richard's rule). The ratio $\mathcal{L}_V/T_b$ or entropy of vaporarization is also roughly constant for metals, but about an order of magnitude larger (Trouton's rule). Using thermochemistry data for iron summarized in Tab.~\ref{Ref:Tab2},   ${C_p}^S (T_f-T_0)+{C_p}^L (T_b-T_f)\sim\mathcal{L}_V/3$ for iron, with $T_0 = 300$~K. Therefore, the energy input required for evaporation of solid Fe is about $4/3\mathcal{L}_V\sim8.25$~MJ/kg. Using laser ablation as a baseline, the fraction of photon energy deposited to the material depends on the laser absorptivity $\chi$, which is about $0.4$ for iron~\cite{Kaplan2014}. Recalling that the input stream is mostly iron, and that the latent energy of vaporization of Nd ($2$~MJ/kg) is about three times smaller than that of Fe, a safe estimate for the laser energy required for the evaporation of NdFeB magnets is $4/3\mathcal{L}_V/\chi\sim20$~MJ/kg. Finally, this value has to be multiplied by the laser electric efficiency $\eta_L$ to provide the true energetic cost of evaporation. Using a poor $\eta_L = 0.1$, this is $200~$MJ/kg.

\begin{table}
\begin{center}
\begin{tabular}{c | c}
 & Fe\\ 
 \hline
Fusion temperature $T_f$ [K] & $1811$\\
Boiling temperature $T_b$ [K]& $3134$\\
Latent heat of fusion $\mathcal{L}_F$ [MJ/kg] & $0.25$\\
Latent heat of vaporization $\mathcal{L}_V$ [MJ/kg] & $6.2$\\
Solid-phase average heat capacity ${C_p}^S$ [kJ/(kg K)] & $0.7$\\
Liquid-phase average heat capacity ${C_p}^L$ [kJ/(kg K)] & $0.82$\\
${C_p}^S (T_f-T_0)+{C_p}^L (T_b-T_f)$  [MJ/kg]& $2.1$\\
\end{tabular}
\caption{Thermochemistry data for iron~\cite{Chase1998}. }
\label{Ref:Tab2}
\end{center}
\end{table}

Moving on to plasma formation and maintenance, an upper bound for the energy cost of plasma formation is obtained by assuming that the plasma formed is fully ionized. Ideally, the input energy required to fully ionize a gas is
\begin{equation}
\mathcal{E}^i = \sum\limits_{j=1}^{\mathcal{N}} {\epsilon^i}_j,
\end{equation}
where $\mathcal{N}$ is the total number of atoms and ${\epsilon^i}_j$ is the ionization energy of atom $j$. Observing that iron atoms (${\epsilon^i}_{\textrm{Fe}} = 7.90$~eV) account for over $80\%$ of all atoms in \ce{Nd2Fe14B} magnets, a first estimate for $\mathcal{E}^i$ is $\mathcal{N}{\epsilon^i}_{\textrm{Fe}} \sim 12$~MJ/kg. This figure is however grossly underestimated since once has to account for all energy dissipation channels. This includes excitation of neutrals atoms and ions, radiation losses, ion heating, etc. For helicon discharges envisioned for plasma filters~\cite{Gueroult2016}, energy dissipation channels have been measured to represent $1.5$ the ionization energy in pure argon plasmas~\cite{Lieberman1994}, or, in other words, a plasma efficiency $\eta_P\sim0.4$.  For more complex plasmas, it stands to reason that this efficiency will be lower. For a mediocre $\eta_P = 0.05$, the cost of plasma formation and maintenance will be about $600$~MJ/kg. Note however that this cost could in principle be significantly lowered if using a partially ionized plasma as discussed in the next paragraph. 

Summing up the cost of feed evaporation and plasma production and maintenance, a rough estimate of the energy cost for plasma processing is  $\mathcal{C}_p \sim 800$~MJ/kg. Assuming an electricity cost of $\$0.12$ per kW.h, this is $\$27$ per kg of magnet. It is important to note here that this cost could be decreased by a factor four since cost targets for solar energy are $\$0.03$ per kW.h by 2030~\cite{SunShot2016}. On the other hand, this derivation of the ionization cost underlines why plasma separation is \emph{a priori} more attractive for concentrated feeds than for diluted feeds. Indeed, since this cost scales with the number of atoms in the feed, the cost of plasma  processing per unit mass of recovered REE will roughly grow linearly with the feed dilution (total number of atoms divided by the number of REEs atoms).  For this reason, plasma techniques appear more promising for recycling large magnets rather than recycling electronic waste. For the same reason, the use of a background gas (\emph{e.~g.} argon), if needed, would increase the processing cost.

A full picture for REEs plasma recycling cost should also include a study of capital, operation and maintenance costs associated with plasma processing. However, our processing cost assessment only dealt with a generic plasma filter, whereas this full cost assessment will require selecting \emph{a priori} a specific plasma filter configuration. Such a detailed analysis is therefore beyond the scope of this work. Nevertheless, to the extent that plasma filtering techniques should have a limited footprint compared to chemical processing, capital costs are expected to compare favourably with those of chemical techniques. Such a favorable cost scaling has for example been suggested when comparing pyro-processing to aqueous techniques for spent fuel reprocessing~\cite{NationalResearchCouncil2001}.

\subsection{Opportunities and perspectives}
\label{Sec:VI}

Mixed REEs used to represent most of the market through their use in catalysts, ceramics and the glass making industry~\cite{Bian2016,Handbook2009}. However, purified REEs presently make for most of the demand owing to their use in fast growing applications ranging from consumer electronics display panels to high-strength magnets. In light of this demand evolution, there is a strong incentive to further process the mixed REEs stream produced by plasma mass separation processes. One option would be to rely on demonstrated chemical processes very similar to those used for REEs extraction from ores. Another option, discussed below, is to carry out this process in a plasma. Among the foreseen advantages of plasmas is the lack of secondary waste stream typically associated with the use of chemicals.    

Separation of individual REEs can be envisioned in what is referred to as advanced separation in the plasma separation flowchart depicted in Fig.~\ref{Fig:FlowChart}. The output of the first stage, made of high purity mixed REEs, is fed into the second stage, where it is separated based on atomic mass into individual REEs. Since plasma mass separation operates by splitting the input feed into two components based on atomic mass (three different components in some cases~\cite{Bardakov2010}), it would take $n\sim\ln{N}/\ln{2}$ stages to separate $N$ distinct REEs. Note that $n\leq5$ since there are $17$ REEs. A non-optimized scheme where particles are collected and then re-ionized while passing from one stage to the next would thus have a cost of about $(1+n)\mathcal{C}_p$, with $\mathcal{C}_p$ the plasma processing cost obtained in the previous section. However, an optimized scheme where ions pass from one stage to the next, with no need for re-ionization, could bring the cost down closer to $\mathcal{C}_p$. This could in principle be achieved in configurations where ions are extracted along the field lines (see, \emph{e.~g.}, Refs.~\cite{Timofeev2007,Fetterman2011,Gueroult2014,Babichev2014}). In the limit of a perfectly optimized scheme, by which it is meant that extra losses induced by the multi-stage scheme are negligible compared to $\mathcal{C}_p$, the cost for recovery of individual REEs would be about $\mathcal{C}_p$.

For a perfectly optimized configuration, producing $1$~kg of Nd from \ce{NdFeB} magnets with $z_{\textrm{Nd}} = 0.27$ would cost $\mathcal{C}_p/z_{\textrm{Nd}}\sim\$99$. To put things into perspective, the market price of $99.5\%$ pure neodymium and dysprosium was respectively $\$39$ and $\$185$ per kg in 2016~\cite{USGS2017}. Due to the larger cost of dysprosium, processing streams with larger dysprosium mass fraction would be more profitable. For example, plasma processing of $1$~kg of a magnet such that $z_{\textrm{Nd}} = 0.25$ and $z_{\textrm{Dy}} = 0.04$ would in principle have a market value of $\$17$, and would cost $\$27$. This plasma processing cost could be brought down to under $\$7$ by 2030 assuming the cost targets for solar energy are met~\cite{SunShot2016}.

\begin{figure}
\includegraphics{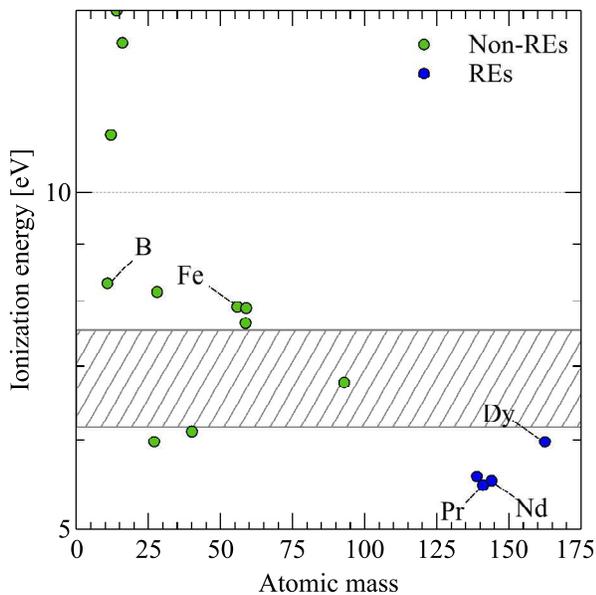}
\caption{Ionization energy of the various elements typically found in Rare Earth permanent magnet waste. Among elements making over $95\%$ of the mass content (Nd, Pr, Dy, Fe and B), a gap in ionization energy of about $2$~eV exists between REEs (in blue) and non-REEs (in light green), as illustrated by the hatched region.}
\label{Fig:Ionization_energy}
\end{figure}

Another promising prospect for \ce{NdFeB} magnets plasma processing lies in the distribution of ionization energy among the various elements found in REMs. Looking at Fig.~\ref{Fig:Ionization_energy}, REEs are found to be the elements with lowest ionization energy ($\leq 6$~eV), while iron, boron and most of the additives and contaminants have ionization energy above $7.5$~eV.  A consequence of this split is that, for given plasma conditions, the ionization fraction of non-REEs should be lower than the ionization fraction of REEs. In plasma processing, the cost is directly proportional to the number of atoms ionized (neglecting evaporation cost). If one could design a scheme such that REEs are fully ionized and hence controllable through the electric and magnetic fields  while a significant fraction of non-REEs (mostly Fe) are not ionized, the cost could be greatly lowered. Quantitatively, assuming only $40\%$ of Fe and B atoms are ionized, the processing cost could be brought down to $\$13$ per kg of \ce{Nd2Fe14B} magnets. Going back to our previous example, 
plasma processing of $1$~kg of a magnet such that $z_{\textrm{Nd}} = 0.25$ and $z_{\textrm{Dy}} = 0.04$ would in principle have a market value of $\$17$, and would cost $\$13$. In other words, it would be profitable. It is worth noting here that benefits of partial ionization could obviously be combined with lower cost for solar energy~\cite{SunShot2016} to further decrease plasma processing costs . 

Not overlooking that this last result is at best a first rough estimate of what the cost of \ce{NdFeB} magnets could be, it commands comments.  
REEs recycling is generally not deemed economically viable owing to recycling processes cost and current REEs prices. However, it has been suggested that an increase of REEs prices by a factor $1.5-10$ could make recycling profitable~\cite{Tsamis2014}, and that technology development should therefore be continued to respond to this eventuality. To the extent that the rough cost estimate derived above is not so high compared to the market value of these elements, and that energy cost for solar could become cheaper~\cite{SunShot2016}, it stands to reason that plasma separation should be further investigated along with other promising processes. The rationale for investigating plasma mass separation for REEs recycling is confirmed when considering that another hurdle on the way to the development of REEs recycling processes is environmental impact. Indeed, since plasma separation does not produce secondary waste (liquid or solid) nor requires chemicals, it should have a very minimal environmental footprint.

\section{Summary and conclusions}
\label{Sec:VII}

In this paper, the potential of plasma mass separation for rare earth recycling was exposed through the example of \ce{NdFeB} magnets recycling. 

Owing to their key role in many fast growing applications, there is presently a clear effort towards securing new supply options for rare earth elements (REEs). One component of supply diversification strategies is recycling. Although chemical techniques used for REEs extraction from primary ores can also be used for this purpose, their environmental impact is a concern. In contrast, plasma separation techniques, which do not require chemicals, have a very minimal environmental footprint.

By analyzing the elemental mass distribution in typical \ce{NdFeB} magnets, it is shown that plasma mass filters have the potential to separate efficiently REEs not only from Fe and B, but also from contaminants an additives. A simple elemental mass separation function, which represents an ideal plasma mass filter, is introduced to quantify separation efficiency.  This model shows that the purity level and extraction efficiency for the separated mixed REEs stream can compare favourably to those of state of the art chemical and high-temperature processes. This simple separation model is further used to identify key parameters controlling the separation efficiency.

A rough estimate of \ce{NdFeB} magnets plasma processing obtained on the basis of the energetic cost of both evaporation and plasma formation and maintenance is shown to be within an order of magnitude of the market value of the recovered mixed REEs. In light of this finding, a multi-stage plasma separation concept allowing to separate individual REEs from one another is presented. This advanced separation concept is finally shown to have the potential to further lower the cost of plasma processing by maximizing the market value of separated elements.

The rough calculations presented here indicate that plasma separation techniques could in principle be implemented to recover rare earth elements from \ce{NdFeB} magnets at a competitive cost and with minimal environmental impact. However, to the extent that large throughput plasma mass filters are still in a relatively primitive stage of development, various elements remains to be demonstrated to confirm these conclusions. In particular, it remains to assess how close practical separation performances can be from the ideal properties used in this study. Nevertheless, in light of the large upside potential of these concepts, it stands to reason that plasma separation should be studied along with advanced chemical and high-temperature techniques.


\section*{Acknowledgements}
This work was supported, in part, by the U.S. DOE OFES Grant $\#$ DE-SC0016072.

\section*{References}

\providecommand{\newblock}{}

\end{document}